\documentclass{mn2e}

\usepackage{graphicx} 
\usepackage{amsmath} 
\usepackage{amssymb} 
\usepackage{natbib} 
\usepackage{bibentry} 
\usepackage{subfig} 
\usepackage{float} 
\usepackage{lscape} 
\usepackage{booktabs} 
\usepackage{color} 
\usepackage[above,below]{placeins} 
\usepackage{appendix}

\usepackage{rotating} 
\usepackage{titlesec} 
\titleformat{\section}{\normalfont\bfseries\raggedright}{\thesection}{1em}{\uppercase}
\titleformat{\subsection}{\normalfont\bfseries\raggedright}{\thesubsection}{1em}{}
\titleformat{\subsubsection}{\normalfont\itshape\raggedright}{\thesubsubsection}{1em}{}
\titlespacing*{\section}{0pt}{25pt}{5pt}
\titlespacing*{\chapter}{0pt}{50pt}{40pt}
\titlespacing*{\section}{0pt}{3.5ex plus 1ex minus .2ex}{2.3ex plus .2ex}
\titlespacing*{\subsection}{0pt}{3.25ex plus 1ex minus .2ex}{1.5ex plus .2ex}
\titlespacing*{\subsubsection}{0pt}{3.25ex plus 1ex minus .2ex}{1.5ex plus .2ex}
\titlespacing{\paragraph}{0pt}{3.25ex plus 1ex minus .2ex}{1em}
\titlespacing{\subparagraph}{\parindent}{3.25ex plus 1ex minus .2ex}{1em}
\renewcommand{\thefootnote}{\fnsymbol{footnote}}

\usepackage[colorlinks=true]{hyperref} 
\definecolor{teal}{rgb}{0.0,0.6,0.4}
\definecolor{maroon}{rgb}{0.8,0.0,0.0}
\definecolor{purple}{rgb}{0.6,0.0,0.6}
\hypersetup{colorlinks, citecolor=blue, linkcolor=maroon, urlcolor=purple}

\usepackage{tabularx} 
\newcolumntype{Y}{>{\centering\arraybackslash}X} 
\newcolumntype{Z}{>{\raggedleft\arraybackslash}X} 

\newcommand{\hersc}{{\it Herschel}}
\newcommand{\planck}{{\it Planck}}
\newcommand{\Kband}{{\it$K_{S}$}-band}

\newcommand{\HI}{H{\sc i}}
\newcommand{\kappad}{$\kappa_{d}$}
\newcommand{\epsilond}{$\varepsilon_{d}$}
\newcommand{\logOH}{$[ 12 + {\rm log}_{10} \frac{O}{H} ]$}
\newcommand{\James}{\citeauthor{James2002}}

\begin{document}

\title[An Empirical Determination of \kappad]{An Empirical Determination of the Dust Mass Absorption Coefficient, \kappad, Using the Herschel Reference Survey}

\author[C.J.R. Clark et\,al.]{Christopher\,J.\,R.\,Clark$^{1\star}$, 
Simon\,P.\,Schofield$^{1}$, 
Haley\,L.\,Gomez$^{1}$,
Jonathan\,I.\,Davies$^{1}$\\
\\
{\parbox{\textwidth}{$^{1}$ School of Physics \& Astronomy, Cardiff University, Queens Buildings, The Parade,  Cardiff, CF24 3AA, UK\\
$^{\star}$ {\tt Christopher.Clark@astro.cf.ac.uk}}}
\\
\\
{\parbox{\textwidth}{\rm {\normalsize Accepted for publication in Monthly Notices of the Royal Astronomical Society}}}}

\date{}

\pagerange{\pageref{firstpage}--\pageref{lastpage}} \pubyear{2015}

\maketitle 

\begin{abstract} We use the published photometry and spectroscopy of 22 galaxies in the \hersc\ Reference Survey to determine that the value of the dust mass absorption coefficient \kappad\ at a wavelength of 500\,\micron\ is $\kappa_{500} = 0.051^{+0.070}_{-0.026}\,{\rm m^{2}\,kg^{-1}}$. We do so by taking advantage of the fact that the dust-to-metals ratio in the interstellar medium of galaxies appears to be constant. We argue that our value for \kappad\ supersedes that of \citet{James2002} -- who pioneered this approach for determining \kappad\ -- because we take advantage of superior data, and account for a number of significant systematic effects that they did not consider. We comprehensively incorporate all methodological and observational contributions to establish the uncertainty on our value, which represents a marked improvement on the oft-quoted `order-of-magnitude' uncertainty on \kappad. We find no evidence that the value of \kappad\ differs significantly between galaxies, or that it correlates with any other measured or derived galaxy properties. We note, however, that the availability of data limits our sample to relatively massive ($10^{9.7}<{M_{\star}}<10^{11.0}\,{\rm M_{\odot}}$), high metallicity (8.61\,\textless\,\logOH\,\textless\,8.86) galaxies; future work will allow us to investigate a wider range of systems.\\
\end{abstract}


\setcounter{footnote}{0}
\renewcommand{\thefootnote}{\textsuperscript{\arabic{footnote}}}

\section{Introduction} \label{Section:Introduction}

The study of cosmic dust has advanced enormously over the past 10--15 years, with the advent of telescopes such as {\it Spitzer} \citep{Werner2004B}, \hersc\ \citep{Pilbratt2010D}, \planck\ \citep{Planck2011I} and ALMA (the Atacama Large Millimetre/submillimetre Array). Observations of dust emission in the Far-InfraRed (FIR) and submillimetre (submm) now serve as some of our most potent tools for understanding the InterStellar Medium (ISM), providing us with avenues to investigate galaxies' chemical evolution, star-formation, and interstellar environments. 

However, our ability to use FIR and submm observations to {\it actually measure the mass of dust} in galaxies is notoriously limited. The dust mass absorption coefficient, \kappad\ (sometimes called the dust mass opacity coefficient), describes what mass of dust gives rise to an observed dust luminosity. However, the value of \kappad\ is very poorly constrained, leading to correspondingly large uncertainty on derived dust mass values. The value of \kappad\ is dictated by the physical properties of the dust, such as the mass density of the constituent materials, the efficiency with which they emit, the grain surface-to-volume ratio, and the grain size distribution. 

A wide range of values of \kappad\ have been estimated, using a variety of techniques. Most require making assumptions about the physical properties of dust grains. The raw materials that make up dust are actually quite well known; the majority of the mass of dust consists of C, N O, Mg, Si, and Fe. This is inferred from observations of the gas phase of the ISM, which is found to be highly depleted of these elements \citep{Savage1996D,Jenkins2009B}. Similarly, some information about the grain size distribution can be extracted from the UltraViolet (UV) dust extinction curve \citep{Kim1994B,Jones1996C,Gall2014A}. Hence chemical considerations can be used to model the mineralogical and physical properties of dust \citep{Whittet1992A,Jones2013A}. Numerous such models exist (eg, \citealp{Hildebrand1983C,Draine1984E,Draine2007A,Jones2013C}), and each implies a corresponding value of \kappad; but there is a great deal of variation between the characteristics of the dust in these various models. Comparisons of FIR/submm emission and UV/optical extinction in Galactic nebul\ae\ can be used to estimate \kappad\ \citep{Casey1991A,Bianchi2003B}, but require assumptions about the cloud geometry, and the results may not apply beyond the nebul\ae\ in question, given the known variation of dust properties with environment \citep{Cardelli1996B,MWLSmith2012B,Planck2013XI,PlanckIntermediateXIV}. A similar approach can be taken with entire nearby galaxies \citep{Alton2000A,Alton2004A,Dasyra2005B}, but this likewise requires assumptions about the geometry and radiative transfer properties of the dust, in order to constrain the optical depth. Laboratory examination of dust analogues, informed by the composition of pre-solar dust grains, provides an alternate approach for determining \kappad\ \citep{Mutschke2013A,Demyk2013A}. However only a small number of truly pre-solar dust grains have been retrieved for analysis \citep{Messenger2013A}, so it is hard to establish the relative importance in the bulk composition of interstellar dust of the particular materials being studied in the laboratory.

The values of \kappad\ suggested by these methods vary enormously. See the summary tables in \citet{Alton2004A} and \citet{Demyk2013A} for a range of observationally- and experimentally-derived values. We can compare values of \kappad\ determined at different wavelengths using the relation:

\begin{equation}
\kappa_{\lambda} = \kappa_{0}\left(\frac{\lambda_{0}}{\lambda}\right)^{\beta}
\label{Equation:Kappa_Wavelength}
\end{equation}

\noindent where $\kappa_{\lambda}$ is the value of \kappad\ at some wavelength $\lambda$, $\kappa_{0}$ is the reference value of \kappad\ at some reference wavelength $\lambda_{0}$, and $\beta$ is the dust emissivity spectral index.

\begin{figure}
\begin{center}
\includegraphics[width=0.495\textwidth]{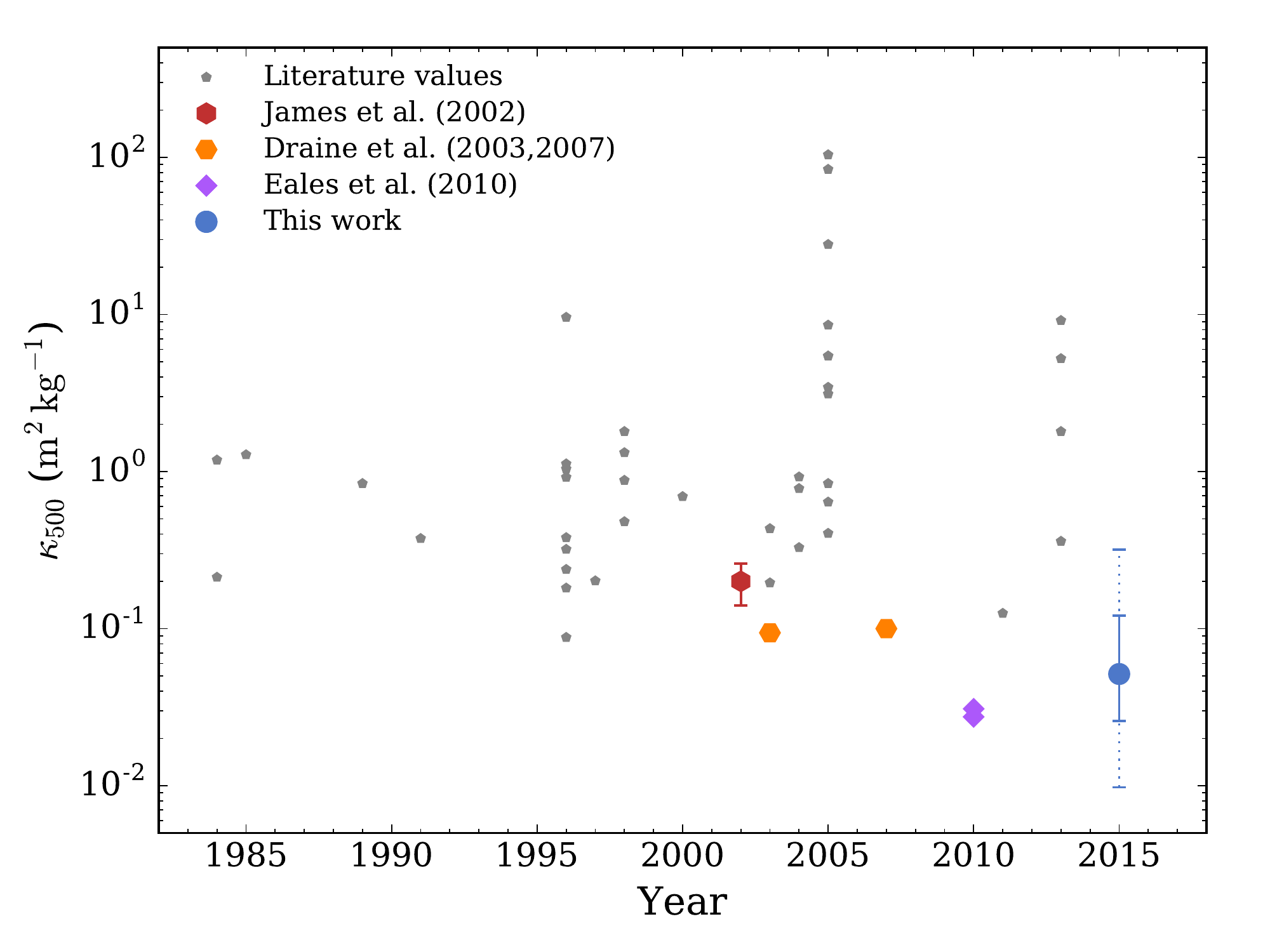}
\caption{Literature values of $\kappa_{500}$, taken from the summary tables of \citet{Alton2004A} and \citet{Demyk2013A}, along with several other widely-cited values \citep{Draine2003A,Dasyra2005B,Draine2007A,Eales2010C,Compiegne2011A}, plotted against their year of publication. There is no indication that reported values of \kappad\ are converging as time goes by. Highlighted is the value of \kappad\ found by \James, whose method this work is based upon; \citet{Eales2010C}, who used a resolved variant of the \James\ method for two galaxies; and the very commonly-used values of \citet{Draine2003A} and \citet{Draine2007A}. Also shown for later comparison is the value of \kappad\ we determine in Section~\ref{Section:Value}. 
The solid error bar shows the uncertainty derived in Section~\ref{Subsection:Uncertainties}, whilst the dotted error bar indicates the potential extent of the systematic offset due to the fact that the absolute metallicity scale calibration is not known to better than 0.7\,dex (\citealp{Kewley2008C}, see Section~\ref{Section:Method}). Note that some of the scatter in this plot will be due to differences in the metallicity prescriptions employed; we opt not to correct for this, therefore keeping the plot representative of the {\it absolute} variation in reported values of \kappad.}
\label{Fig:Year_vs_Kappa}
\end{center}
\end{figure}

The literature values of \kappad\ listed in the summary tables of \citet{Alton2004A} and \citet{Demyk2013A}, along with several other commonly-cited values \citep{Draine2003A,Dasyra2005B,Draine2007A,Eales2010C,Compiegne2011A}, are plotted in Figure~\ref{Fig:Year_vs_Kappa}, converted\footnote{Where the \citet{Demyk2013A} summary table states a \kappad\ that depends upon temperature, with entries for both 300\,K and 10\,K, the 10\,K value has been taken.} to $\kappa_{500}$ as per Equation~\ref{Equation:Kappa_Wavelength} (only using values where $\lambda_{0}\,\geq\,250$\,\micron, and assuming $\beta = 2$ as a basic approximation). These 46 values have a standard deviation of 0.8\,dex, and span {\it over 3.5 orders of magnitude} in total, ranging from $\kappa_{500} = 0.031\,{\rm m^{2}\,kg^{-1}}$ to $\kappa_{500} = 104\,{\rm m^{2}\,kg^{-1}}$. 

This vast uncertainty in \kappad\ is extremely troubling, especially considering that the observed dust mass of a galaxy is now being used as a proxy for estimating other quantities, such as the total gas mass \citep{Eales2012A,Scoville2014B}. Modulo the uncertainty on the value of \kappad, this promises to be a useful tool, particularly at high redshift where other gas estimators may not be available.

Ideally, in order to calibrate a robust value for \kappad, there would be a way to {\it a priori} know the dust mass present in a galaxy, without reference to FIR/submm observations. Fortunately, \citet{James2002} demonstrated that this is, in fact, possible. It has been 13 years since \James\ first applied their technique; with the advent of \hersc, and the greatly improved quality of extragalactic observations now available, the time is now ripe to repeat their analysis, taking advantage of the the greatly improved resources at our disposal.

In Section~\ref{Section:Method} we describe the \James\ method, and how we intend to apply it. In Section~\ref{Section:Sample} we outline the \hersc\ Reference Survey, the sample we use to perform our analysis. In Section~\ref{Section:SED_Fitting} we describe how we fit the dust SEDs of the galaxies in our sample. In Section~\ref{Section:Value} we put the method into practice, to arrive at a new, well-constrained value for \kappad. In Section~\ref{Section:Variation} we look at how our computed values for \kappad\ vary across our sample. In Section~\ref{Section:Comparison} we compare our value for \kappad\ to other reported values.

\section{The Method} \label{Section:Method}

The \James\ method for determining the value of \kappad\ takes advantage of the fact that the fraction of the metals in a galaxy's ISM that are locked up in dust, \epsilond, appears to be constant. A wealth of evidence supports the notion that \epsilond\ is constant in the modern universe \citep{Sodroski1997A,Dwek1998B,Leroy2011B,Watson2011A,DJBSmith2012,Corbelli2012A}, with further work suggesting it is also constant at high redshift \citep{Pei1992A,Pei1999A,Zafar2013D,Chen2013B,Sparre2014A}. There also exist theoretical frameworks to explain the observed invariance in \epsilond\ \citep{Inoue2003C,Asano2011A,Mattsson2014A}. It should however be noted that there is conflict in the literature on the matter of whether or not \epsilond\ remains constant in low-metallicity systems; some studies find that \epsilond\ differs in low-metallicity dwarf galaxies \citep{Galliano2005B,Hunt2005A}, whereas others have found \epsilond\ to be constant over a wide range of masses, metallicities, and redshifts \citep{Pei1992A,Zafar2013D}. 

Many of the interstellar dust-to-metals ratios quoted in the literature are not suitable for the purposes of this work.  Some quote the dust-to-metals ratio in terms of extinction per column density of metals, $A_{V} / N_{H_{X}}$ (eg, \citealp{Watson2011A,Zafar2013D,Sparre2014A}), which cannot be converted to a value of \epsilond\ without assuming what column density of dust corresponds to a given degree of extinction -- and given that the aim of this work is the determine the dust mass-to-emission ratio, it would be unwise to predicate it upon assumptions about the equally poorly-constrained dust mass-to-extinction ratio. There are also works which use an assumed value of \kappad\ to arrive at their value for \epsilond\ (eg, \citealp{DJBSmith2012,Davies2014A}); as such, any attempt to use these values to estimate \kappad\ would be an exercise in circular logic.

Fortunately, there are numerous values of \epsilond\ reported in the literature which are suitable for the purposes of determining \kappad, from studies which examine elemental depletions to establish the fraction of metals locked up in dust grains. The values we consider are 0.5 \citep{Issa1990B}, 0.36 (\citealp{Luck1992A}, Large Magellanic Cloud), 0.46 (\citealp{Luck1992A}, Small Magellanic Cloud), 0.5 \citep{Whittet1992A}, 0.51 \citep{Pei1992A} 0.36 \citep{Meyer1998B}, 0.3 \citep{Dwek1998B}, 0.45 \citep{Pei1999A}, 0.529 \citep{Weingartner2001A}, 0.456 \citep{James2002}, 0.549 \citep{Kimura2003C}, and 0.387 \citep{Draine2007C}. The average (mean, median, and mode) of these 12 values is 0.5, and the standard deviation is 0.1; hence we adopt a value for the interstellar dust-to-metals ratio of $\varepsilon_{d} = 0.5 \pm 0.1$. 

Using \epsilond\ requires knowledge of the metallicity of a galaxy's ISM. Gas-phase metallicity is generally expressed in terms of the logarithm of the oxygen-to-hydrogen bulk abundance ratio, in the form \logOH; in such units, the Solar metallicity is $8.69 \pm 0.05$ \citep{Asplund2009A}. Because we are concerned with the absolute metal mass fraction, we convert \logOH\ metallicites to metal mass fractions by reference to the Solar values; we use a Solar metal mass fraction of $f_{Z_{\odot}} = 0.0134$ (\citealp{Asplund2009A}\footnote{We note that this value for the Solar metal mass fraction is $\sim$\,33\,per\,cent lower than values typically used pre-2005 (see \citealp{Asplund2009A}).}, uncertainty assumed to be negligible). This does entail the assumption that \logOH\ metallicity is a direct proxy for absolute metallicity. A wide variety of combinations of atomic species and emission lines are commonly used to estimate \logOH\ metallicity (O{\sc ii}, O{\sc iii}, N{\sc ii}, S{\sc ii} H$\alpha$, H$\beta$, etc); \citet{Kewley2008C} have shown how different metallicity prescriptions can be normalised to give \logOH\ values with a relative accuracy of \textless\,0.1\,dex, but that the {\it absolute} metallicity scale calibration is not known to better than 0.7\,dex. The represents a systematic uncertainty on any absolute metallicity, including those we use here. Because it is a systematic, we do not incorporate it into our uncertainty analysis in Section~\ref{Subsection:Uncertainties}; rather, we stress to the reader that there is some underlying fixed offset between the metal mass fractions we (and any other authors) employ, and the corresponding true metal mass fractions, of up to 0.7\,dex.

It is also important to note that \logOH\ is a tracer of {\it gas}-phase metallicity -- whereas we are interested in the metallicity of the entire ISM. When using gas-phase oxygen abundance to determine ISM metallicity, a correction needs to be applied to account for the the fraction of oxygen depleted onto dust grains. Specifically, this correction needs to account for the oxygen depletion level in H{\sc ii} regions, which are the source of the majority of the nebular line emission of star-forming galaxies \citep{Kunth2000A}, and hence are where empirical gas-phase metallicity estimators are calibrated (including those used in this work, see Section~\ref{Section:Sample}). We perform this correction using the \citet{Mesa-Delgado2009B} oxygen-depletion factor of $\delta_{O} = 1.32 \pm 0.09$, determined by observing reductions in oxygen depletion due to dust destruction in shocked regions of Orion Nebula. This value is in good agreement with the values of $\delta_{O}$ reported by \citet{Zurita2012A} from comparisons of H{\sc ii} regions and blue supergiants in M\,31, \citet{Patterson2012C} and \citet{Kudritzki2012B} for M\,81, and \citet{Peimbert2012C} for galactic H{\sc ii} regions. It should be noted that studies suggests $\delta{_O}$ decreases in low-metallicity environments, with no depletion observed by \citet{Bresolin2009B} in NGC\,0300 (\logOH\,$\sim$\,8.3), and $\delta_{O} = 1.20$ found by \citet{Peimbert2012C} in low-metallicity galaxy SBS 0335-052 E (\logOH\,$\sim$\,7.4). However, as all of the galaxies we consider in this work have at least double these metallicities (\logOH\,\textgreater\,8.61, see Section~\ref{Section:Sample}), the value of $\delta_{O}$ we adopt should be unaffected.

Assuming that \epsilond\ is indeed constant, the dust mass $M_{d}$ in a galaxy will be given by:

\begin{equation}
M_{d} =  M_{g}\, \varepsilon_{d}\, f_{Z_{\odot}}\, Z
\label{Equation:James_Dust}
\end{equation}

\noindent where $M_{g}$ is the gas mass of a galaxy's ISM, $f_{Z_{\odot}}$ is the metal mass fraction at Solar metallicity, and $Z$ is the metallicity of a galaxy's ISM as a fraction of the Solar value. 

The value of $Z$ for the galaxy in question is arrived at using its $\left[\frac{O}{H}\right]$ bulk abundance ratio measurement (converted from \logOH\ format), corrected for gas-phase oxygen depletion, according to:

\begin{equation}
Z =\delta_{O} \left[\frac{O}{H}\right] \bigg/ \left[\frac{O}{H}\right]_{\odot}
\label{Equation:Metallicity}
\end{equation}

\noindent where $[\frac{O}{H}]_{\odot}$ is the Solar oxygen bulk abundance ratio.

The total mass of a galaxy's ISM is:

\begin{equation}
M_{g} = \xi ( M_{{\it H{\sc I}}} + M_{H_{2}} )
\label{Equation:Total_Gas}
\end{equation}

\noindent where $M_{{\it H{\sc I}}}$ is the mass of atomic hydrogen, $M_{H_{2}}$ is the mass of molecular hydrogen, and $\xi$ is a correction factor to account for the fraction of a galaxy's ISM made up of elements heavier than hydrogen, defined as:

\begin{equation}
\xi =  \frac{ 1 }{ 1 - \left(f_{\it He_{p}} + f_{Z} \left[\frac{\Delta f_{\it He}}{\Delta f_{Z}}\right] \right) - f_{Z} }
\label{Equation:xi}
\end{equation}

\noindent where $f_{\it He_{p}}$ is the primordial helium mass fraction of $0.2485 \pm 0.0002$ \citep{Aver2013A}, $f_{Z}$ is the metal mass fraction of the galaxy in question (such that $f_{Z} = f_{Z_{\odot}} Z$), and $[\frac{\Delta f_{\it He}}{\Delta f_{Z}}]$ is the evolution of the helium mass fraction with metallicity (such that the helium mass fraction $f_{\it He} = f_{\it He_{p}} + f_{Z} [\frac{\Delta f_{\it He}}{\Delta f_{Z}}]$) for which we use a rate of $[\frac{\Delta f_{\it He}}{\Delta f_{Z}}] = 1.41 \pm 0.62$ \citep{Balser2006D}. In the literature, it is common to account for helium alone, and assume solar metallicity when doing so, equivalent to using a correction factor of $\xi = 1.36$. However, this ignores the mass contribution of metals, and the fact that the helium fraction is metallicity-dependant; a galaxy with zero metallicity would have $\xi = 1.33$, whilst $\xi = 1.39$ at solar metallicity.

The mass of atomic hydrogen (in Solar masses) in a galaxy is determined from 21\,cm observations using the standard prescription:

\begin{equation}
M_{{\it H{\sc I}}} =  2.356\times10^{5} S_{\it H{\sc I}}\, D^{2}
\label{Equation:Atomic_Hydrogen}
\end{equation}

\noindent where $S_{\it H{\sc I}}$ is the integrated 21\,cm line flux density in Jy\,km\,s$^{-1}$, and $D$ is the source distance in Mpc.  

The mass of molecular hydrogen in a galaxy (in Solar masses) is typically inferred from the luminosity of the $^{12}$C$^{16}$O(1-0) line, according to:

\begin{equation}
M_{H_{2}} = 2.453\times10^{3}\, S_{\it CO}\, D^{2}\, \alpha_{\it CO}
\label{Equation:Molecular_Hydrogen}
\end{equation}

\noindent where $S_{\it CO}$ is the integrated flux density of the $^{12}$C$^{16}$O(1-0) line in Jy\,km\,s$^{-1}$, $D$ is the source distance in Mpc, and $\alpha_{\it CO}$ is the CO-to-H$_{2}$ conversion factor in M$_{\odot}$\,K$^{-1}$\,km$^{-1}$\,s. We opt to use the metallicity-dependant $\alpha_{\it CO}$ prescription of \citet{Schruba2012B}, which takes the form:

\begin{equation}
{\rm log}_{10} (\alpha_{\it CO}\,\xi) =  {\rm log}_{10}(A) + N \left( \left[ 12 + {\rm log}_{10} \frac{O}{H} \right] -8.7  \right)
\label{Equation:Schruba_Alpha}
\end{equation}

\noindent where $A$ and $N$ are empirical calibration constants determined by \citet{Schruba2012B} with values $A = 8.0 \pm 1.3$ and $N = -2.0 \pm 0.4$. \citet{Schruba2012B} calibrated this prescription empirically over a 1\,dex range in metallicity (8\,\textless\,\logOH\,\textless\,9), encompassing the full metallicity range of the galaxies we consider in this work (see Section~\ref{Section:Sample}), and find 0.1\,dex of scatter on the prescription as a whole. Note that for the metallicity range of of the galaxies we consider in this work (see Section~\ref{Section:Sample}), the \citet{Schruba2012B} prescription is compatible with the alternative metallicity-dependant prescription of \citet{Genzel2012A}.

We include the $\xi$ term in Equation~\ref{Equation:Schruba_Alpha} to account for the fact that the \citet{Schruba2012B} prescription is calibrated using the rate at which star-forming material is consumed, and hence includes the mass of helium (and other elements) associated with the molecular hydrogen being traced by the CO (A.\,K.\,Leroy, {\it priv. comm.}) -- whereas we are only concerned with the mass of molecular hydrogen. 

The FIR--submm (50\,\textless\,$\lambda$\,\textless\,1000\,\micron) emission from dust in a galaxy is described by a two-component modified blackbody Spectral Energy Distribution (SED), which takes the form:

\begin{equation}
M_{d} = \frac{ S_{\lambda_{w}}\, D^{2} }{ \kappa_{\lambda}\, B_{\lambda}(T_{w}) } + \frac{ S_{\lambda_{c}}\, D^{2} }{ \kappa_{\lambda}\, B_{\lambda}(T_{c}) }
\label{Equation:SED}
\end{equation}

\noindent where $S_{\lambda_{w}}$ and $S_{\lambda_{c}}$ are the flux densities of the warm and cold dust components at wavelength $\lambda$ in W\,Hz\,$^{-1}$\,m$^{-2}$, and $B_{\lambda}(T_{w})$ and $B_{\lambda}(T_{c})$ are each the Planck function at wavelength $\lambda$ and characteristic dust temperatures $T_{w}$ and $T_{c}$. Whilst a single-component modified blackbody would be a simpler model, recent work has shown that this approach can systematically fail to fit the SEDs of certain galaxies; we expand upon this, and detail how our SED fitting is performed, in Section~\ref{Section:SED_Fitting}.

Substituting Equation~\ref{Equation:Total_Gas} into Equation~\ref{Equation:James_Dust}, setting that equal to Equation~\ref{Equation:SED}, and re-arranging to make $\kappa_{\lambda}$ the subject, gives us the formula:

\begin{equation}
\kappa_{\lambda} = \frac{ D^{2} }{ \xi\, ( M_{{\it H{\sc I}}} + M_{H_{2}} )\, \varepsilon_{d}\, f_{Z_{\odot}}\, Z } \left( \frac{ S_{\lambda_{w}} }{ B_{\lambda}(T_{w}) } + \frac{ S_{\lambda_{c}} }{ B_{\lambda}(T_{c}) } \right)
\label{Equation:Kappa}
\end{equation}

\noindent which can be used to empirically determine the value of $\kappa_{d}$ for any galaxy for which FIR--submm photometry, atomic gas mass, molecular gas mass, and integrated gas-phase metallicity is available. Note that the resulting value of \kappad\ is not affected by uncertainty in source distance, because the terms for both $M_{{\it H{\sc I}}}$ and $M_{H_{2}}$ are proportional to $D^{2}$; as such $D^{2}$ ultimately cancels out of Equation~\ref{Equation:Kappa}.

Whilst the method we have laid out here follows the same basic principle as that of \James, we note that they did not explicitly account for the mass helium or metals when considering their ISM masses, whereas we do so by including the $\xi$ term in Equation~\ref{Equation:Total_Gas}; the metallicites of the galaxies we consider in this work (see Section~\ref{Section:Sample}) give a median value of $\xi = 1.41$, and hence our ultimate value of \kappad\ will also be reduced by a factor of 1.41 (see Equation~\ref{Equation:Kappa}). Nor did the \James\ method appear to account for the depletion of oxygen onto dust, despite the fact that they used measurements of gas-phase oxygen abundance, pegged to Solar values, to determine absolute metallicities; given our adopted correction of $\delta_{O} = 1.32$, the factor by which our values of \kappad\ will be reduced as a result of this consideration will likewise be 1.32. Combined, $\xi$ and $\delta_{O}$ will reduce any value of \kappad\ by a factor of 1.86 -- our inclusion of these systematic effects hence represents an essential development of the \James\ technique.

\section{The Sample} \label{Section:Sample}

To perform our determination of \kappad, we use the rich, homogeneous dataset of the \hersc\ Reference Survey (HRS, \citealp{Boselli2010}). The HRS consists of 323 galaxies in the velocity range $1050 \le V \le 1750\ \rm km\ s^{-1}$ (with corrections made to account for the velocity dispersion of the galaxies of the Virgo Cluster), corresponding to a distance range of $15 \le D \le 25\,{\rm Mpc}$. The HRS galaxies were selected on the basis of their \Kband\ brightness, as this is the part of the stellar emission spectrum that suffers least from extinction, and is known to be a good proxy for stellar mass. The apparent magnitude limit of the late type galaxies in HRS is $K_{S} \le 12$, which equates to an absolute magnitude limit in the range $-17.43 \leq\ K_{S} \leq -18.54$, depending on the distance of the source between the HRS limits. For early type galaxies, a brighter apparent magnitude limit of $K_S \le 8.7$ is applied. 

The HRS has excellent multiwavelength photometric data available. The \hersc-SPIRE\footnote{Spectral and Photometric Imaging REceiver \citep{Griffin2010D}} photometry is presented in \citet{Ciesla2012B}, \hersc-PACS\footnote{Photodetector Array Camera and Spectrometer \citep{Poglitsch2010B}} photometry in \citet{Cortese2014A}, WISE\footnote{Wide-field Infrared Survey Explorer \citep{Wright2010F}} photometry in \citet{Ciesla2014A}, and SDSS\footnote{Sloan Digital Sky Survey \citep{York2000B}} and GALEX\footnote{GAlaxy Evolution EXplorer \citep{Morrissey2007B}} photometry in \citet{Cortese2012C} (note that not all of this photometry is used in our determination of \kappad; the SDSS and GALEX data is used in Section~\ref{Section:Variation}, when comparing \kappad\ to other galaxy properties).

Drift-scan spectroscopy of the HRS galaxies is presented in \citet{Boselli2013A}; this data is in turn used by \citet{Hughes2013B} to determine integrated gas-phase metallicities, normalised for direct comparison as per the prescriptions of \citet{Kewley2008C}. \citet{Boselli2014B} present 21\,cm atomic hydrogen observations and $^{12}$C$^{16}$O(1-0) molecular gas observations. We note that most of the CO observations in \citet{Boselli2014B} are single-dish central pointings, with only partial coverage of the target galaxies (they infer the total CO emission assuming exponential molecular gas discs); they do, however, present and homogenise literature CO observations that fully map HRS galaxies.

To use Equation~\ref{Equation:Kappa}, we need FIR--submm photometry (with which to fit the dust SED), 21\,cm measurements, $^{12}$C$^{16}$O(1-0) measurements, and gas-phase metallicities -- all integrated over the entire target galaxy. In total, 22 HRS galaxies have this complete range of data available; these are the galaxies we use in Section~\ref{Section:Value} when determining the value of \kappad. These 22 galaxies span a stellar mass range of $10^{9.7}<{M_{\star}}<10^{11.0}\,{\rm M_{\odot}}$, and a metallicity range of 8.61\,\textless\,\logOH\,\textless\,8.86.


This dataset represents an enormous improvement over what was available to \James\ for their original determination of \kappad. In particular, the \hersc\ photometry of the HRS galaxies allows for far better SED fitting than was possible with the IRAS\footnote{InfraRed Astronomical Satellite \citep{Neugebauer1984}} 12--100\,\micron\ and JCMT SCUBA\footnote{James Clerk Maxwell Telescope Submillimetre Common-User Bolometer Array \citep{Holland1999A}} 850\,\micron\ data that \James\ had at their disposal. Similarly, the metallicities used by \James\ were not integrated measurements of the kind available for the HRS, but instead were derived from observations of a few individual H{\sc ii} regions only, and pre-date the metallicity-normalisation procedure of \citet{Kewley2008C}.

The basic properties of the galaxies in our sample are given in Table~\ref{AppendixTable:Sample}, and the gas masses and metallicities are given in Table~\ref{AppendixTable:Gas}

\section{SED Fitting} \label{Section:SED_Fitting}

We opt to use 500\,\micron\ as our reference wavelength for determining \kappad, as it is a common choice in the literature (allowing for easy comparison), and because it is the longest \hersc\ wavelength, and hence the least affected by dust temperature. 

For each source, we determine $S_{500_{c}}$, $S_{500_{w}}$, $T_{c}$, and $T_{w}$, by fitting a two-component modified blackbody model to the dust SED from 60--500\,\micron, using a $\chi^{2}$-minimising routine which incorporates the colour-corrections for filter response function and beam area\footnote{SPIRE handbook: \url{http://herschel.esac.esa.int/Docs/SPIRE/spire_handbook.pdf}.}\textsuperscript{,}\footnote{PACS instrument and calibration wiki: \url{http://herschel.esac.esa.int/twiki/bin/view/Public/PacsCalibrationWeb}.}\textsuperscript{,}\footnote{IRAS LAMBDA explanatory supplement: \url{http://lambda.gsfc.nasa.gov/product/iras/}}\textsuperscript{,}\footnote{WISE all-sky data relase explanatory supplement: \url{http://wise2.ipac.caltech.edu/docs/release/allsky/expsup/}.}. Note that for a galaxy with an SED that is well-fit by a single-component model, this method is free to assign negligible mass to one of the dust components, or fit two identical-temperature components. We use the 100, 160, 250, 350, and 500\,\micron\ fluxes published by the HRS\footnote{We corrected the HRS fluxes to account for a recently-fixed error in the {\tt Scanamorphos} pipeline \citep{Roussel2013A} used to create the HRS PACS maps. The published HRS fluxes at 100 and 160\,\micron\ were multiplied by 1.01 and 0.93 respectively, the average change (with scatter $\sim$2\,per\,cent) in extended-source flux in maps produced with corrected versions of {\tt Scanamorphos}.}, whilst at 60\,\micron\ we use IRAS photometry obtained using the Scan Processing and Integration Tool (SCANPI\footnote{\url{http://irsa.ipac.caltech.edu/applications/Scanpi/}}), following the procedure laid out by \citet{Sanders2003A}. Note that this photometry is all at sufficiently long wavelengths that it will be unaffected by Polycyclic Aromatic Hydrocarbon (PAH) emission, which occurs at wavelengths $\lessapprox$\,20\micron\, \citep{Draine2007A,DaCunha2008}. We also use the 22\,\micron\ fluxes published by the HRS as upper limits, to prevent unconstrained warm components from being fitted. 

When modelling FIR--submm SEDs, there is a well-established degeneracy between temperature and $\beta$ \citep{Shetty2009A,DJBSmith2013A}, that leads to an artificial anticorrelation. To confuse matters further, methods that should be `immune' to the temperature-$\beta$ degeneracy give conflicting results regarding the actual variation of $\beta$ with temperature; the hierarchical Bayesian fitting approach of \citet{Kelly2012B} indicates that temperature and $\beta$ are positively correlated, whilst the laboratory analysis of \citet{Demyk2013A} suggests that there is in fact a real, underlying temperature-$\beta$ anticorrelation. For these reasons, we opt to employ a fixed $\beta$ in this work; specifically, we use a value of $\beta=2$, as both observational \citep{Dunne2001A,Clemens2013A,DJBSmith2013A,PlanckIntermediateXIV} and experimental \citep{Demyk2013A} evidence suggests that values in the range 1.8--2.0 are appropriate for nearby galaxies. Recent work has shown that when keeping $\beta$ fixed, a single-component modified blackbody SED can systematically fail to fit the dust emission of galaxies (particularly in the case of late types), whilst a two-component model works well \citep{CJRClark2015A}; hence this is the model we use (moreover, \citealp{Remy-Ruyer2015B} have shown that the single-component modified blackbody approach can systematically fail even when $\beta$ is left free to vary). Fixing $\beta$ does, however, artificially reduce the uncertainty in SED fits; we address this in Section~\ref{Subsection:Uncertainties}. 

The SEDs are shown in Figure~\ref{AppendixFig:SED_Grid}. The best-fit values for each parameter are given in Table~\ref{AppendixTable:Dust}.

\begin{table}
\begin{center}
\caption{Values and uncertainties of $\kappa_{500}$ determined for the galaxies of our sample, and the sample as a whole. The $\pm$ uncertainties on each value are asymmetric, and are defined by the $66.\dot{6}$\textsuperscript{th} percentiles away from the determined value along the bootstrapped distributions (in each direction). We also quote approximately-equivalent logarithmic uncertainties (in dex), defined by the $66.\dot{6}$\textsuperscript{th} percentile away from the determined values in absolute terms (ie, in both directions). The overall sample value of $\kappa_{500}$ is the median value.}
\label{Table:Kappa}
\begin{tabular}{lrrrrr}
\toprule \toprule
\multicolumn{1}{c}{Name} &
\multicolumn{1}{c}{$\kappa_{500}$} &
\multicolumn{1}{c}{$-\,\kappa_{500}$} &
\multicolumn{1}{c}{$+\,\kappa_{500}$} &
\multicolumn{1}{c}{$\Delta\,\kappa_{500}$} \\
\cmidrule(lr){2-4}
\multicolumn{1}{c}{} &
\multicolumn{3}{c}{(${\rm m^{2}\,kg^{-1}}$)} &
\multicolumn{1}{c}{(dex)}\\
\midrule
NGC 3437 & 0.048 & -0.039 & +0.078 & 0.157\\ 
NGC 3631 & 0.042 & -0.029 & +0.118 & 0.252\\ 
NGC 3683 & 0.055 & -0.043 & +0.147 & 0.310\\ 
NGC 3953 & 0.064 & -0.045 & +0.186 & 0.285\\ 
NGC 4030 & 0.055 & -0.045 & +0.125 & 0.272\\ 
M 98 & 0.063 & -0.051 & +0.119 & 0.195\\ 
NGC 4212 & 0.059 & -0.047 & +0.132 & 0.245\\ 
M 99 & 0.037 & -0.030 & +0.085 & 0.248\\ 
M 61 & 0.039 & -0.031 & +0.079 & 0.210\\ 
M 100 & 0.067 & -0.054 & +0.146 & 0.243\\ 
M 86 & 0.054 & -0.043 & +0.131 & 0.283\\ 
M 88 & 0.071 & -0.057 & +0.163 & 0.274\\ 
NGC 4527 & 0.042 & -0.035 & +0.084 & 0.211\\ 
NGC 4535 & 0.067 & -0.055 & +0.127 & 0.203\\ 
NGC 4536 & 0.062 & -0.033 & +0.154 & 0.242\\ 
NGC 4567 & 0.014 & -0.011 & +0.038 & 0.335\\ 
NGC 4568 & 0.061 & -0.042 & +0.209 & 0.356\\ 
M 60 & 0.033 & -0.025 & +0.092 & 0.327\\ 
NGC 4651 & 0.042 & -0.033 & +0.077 & 0.179\\ 
NGC 4654 & 0.046 & -0.037 & +0.086 & 0.194\\ 
NGC 4689 & 0.049 & -0.038 & +0.139 & 0.342\\ 
NGC 5248 & 0.042 & -0.035 & +0.080 & 0.203\\ 
\bottomrule
Overall & 0.051 & -0.026 & +0.070 & 0.244\\
\bottomrule
\end{tabular}
\end{center}
\end{table}

\section{Determining The Dust Mass Absorption Coefficient} \label{Section:Value}

We now have the values necessary to use Equation~\ref{Equation:Kappa} to determine $\kappa_{500}$ for each of the galaxies in our sample; the resulting values are listed in Table~\ref{Table:Kappa}. The values range from $\kappa_{500} = 0.031\,{\rm m^{2}\,kg^{-1}}$ (for NGC\,4567), to $\kappa_{500} = 0.071\,{\rm m^{2}\,kg^{-1}}$ (for M\,88). The median value is $\kappa_{500} = 0.051\,{\rm m^{2}\,kg^{-1}}$.

\begin{figure}
\begin{center}
\includegraphics[width=0.485\textwidth]{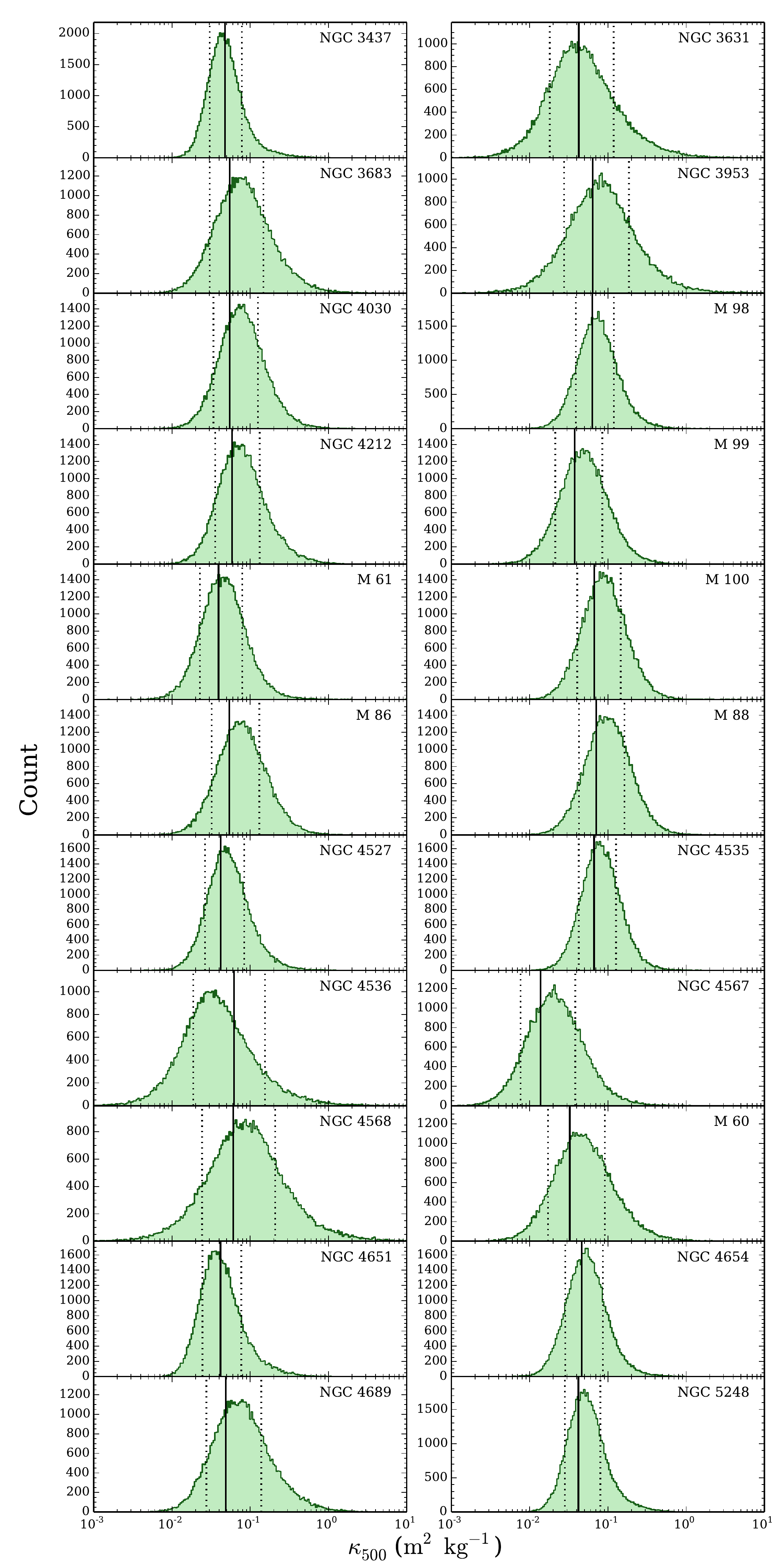}
\caption{The distributions of values of $\kappa_{500}$ produced by bootstrapping Equations~\ref{Equation:Metallicity}, \ref{Equation:xi}, \ref{Equation:Atomic_Hydrogen}, \ref{Equation:Molecular_Hydrogen}, \ref{Equation:Schruba_Alpha}, and \ref{Equation:Kappa}, for each of the galaxies in our sample. The actual value of $\kappa_{500}$ determined for each galaxy is indicated by the solid black line. We define the uncertainties on each value by the $66.\dot{6}$\textsuperscript{th} percentiles away from the determined value along the bootstrapped distributions (in each direction); these are indicated by the dotted black lines.}
\label{Fig:Kappa_Bootstrap_Hist_Grid}
\end{center}
\end{figure}

\begin{figure}
\begin{center}
\includegraphics[width=0.495\textwidth]{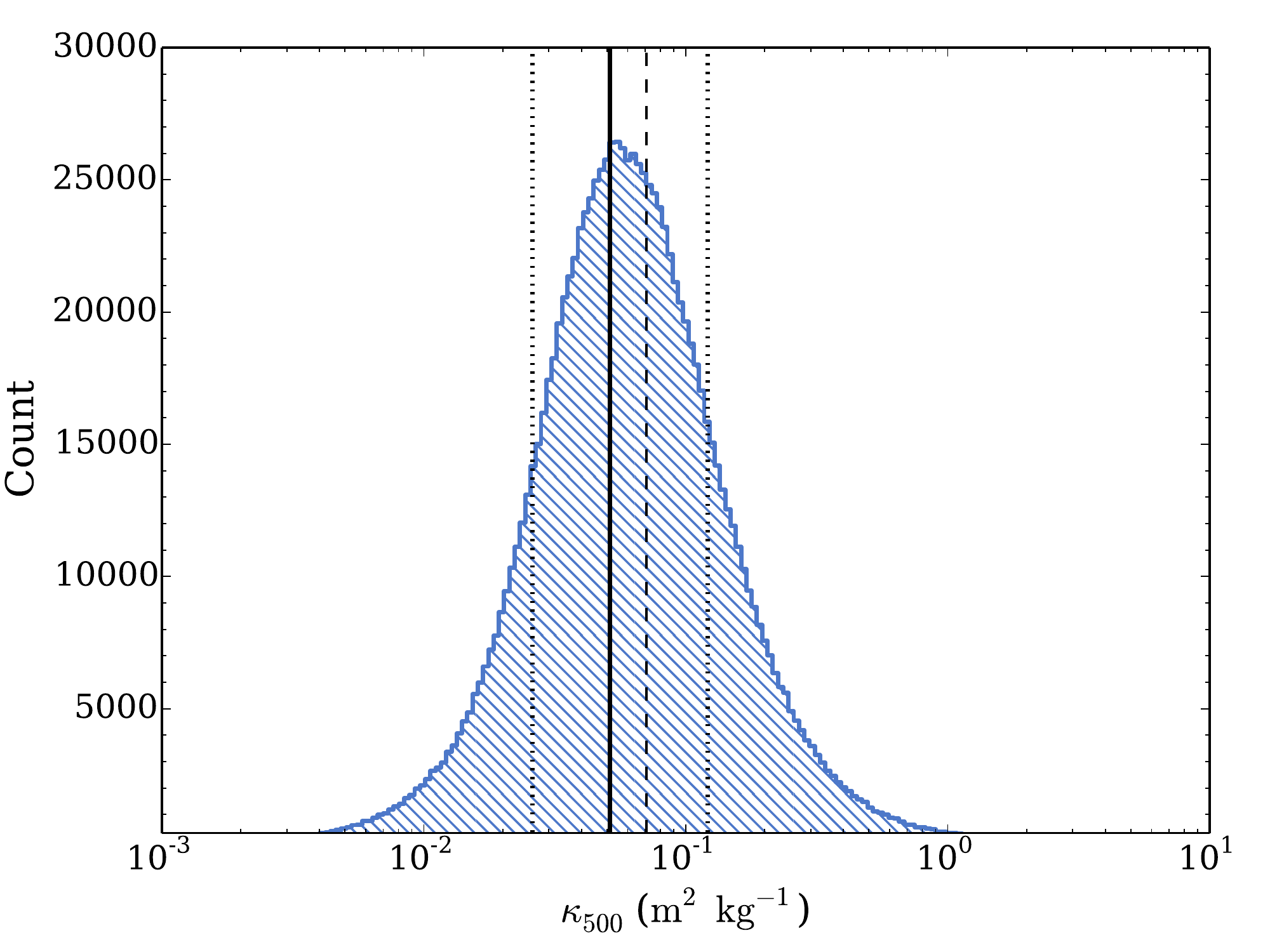}
\caption{The distribution of values of $\kappa_{500}$ produced by combining all of the bootstrapped distributions in Figure~\ref{Fig:Kappa_Bootstrap_Hist_Grid}.  The median determined value of $\kappa_{500}$  is indicated by the solid black line. The dotted black lines indicate the uncertainties on this value, defined by the $66.\dot{6}$\textsuperscript{th} percentiles away from the median value along the bootstrapped distribution (in each direction). Also plotted, for comparison, is a dashed black line indicating the alternate value of $\kappa_{500}$ calculated in Section~\ref{Subsection:Milky_Way_Alpha_CO}, using a constant Milky Way $\alpha_{\it CO}$ (as opposed to the metallicity-dependant value used for the main determination).}
\label{Fig:Kappa_Bootstrap_Hist}
\end{center}
\end{figure}

\subsection{Uncertainties} \label{Subsection:Uncertainties}

To determine the uncertainty on the value of $\kappa_{500}$ for each galaxy in our sample, we employ a Monte Carlo bootstrapping analysis, whereby the parameters in Equations~\ref{Equation:Metallicity}, \ref{Equation:xi}, \ref{Equation:Atomic_Hydrogen}, \ref{Equation:Molecular_Hydrogen}, \ref{Equation:Schruba_Alpha}, and \ref{Equation:Kappa} were re-sampled, and the value of $\kappa_{500}$ was re-calculated; this process was repeated 50,000 times for each source.

We generated re-sampled values of \epsilond, $S_{{\it H{\sc I}}}$, $S_{\it CO}$,  $\delta_{O}$, \logOH\ (and hence $Z$), $[\frac{O}{H}]_{\odot}$, $f_{\it He_{p}}$, $[\frac{\Delta f_{\it He}}{\Delta f_{Z}}]$, $A$, $N$, $\alpha_{\it CO}$\footnote{A re-sampled value of $\alpha_{\it CO}$ is dictated by the perturbed values of \logOH, $A$, and $N$, as per Equation~\ref{Equation:Schruba_Alpha}, generated by randomly perturbing each according to a Gaussian distribution defined by their given uncertainties. However, as previously stated, \citet{Schruba2012B} find 0.1\,dex of scatter on their prescription as a whole; we therefore further perturbed each generated value of $\alpha_{\it CO}$ accordingly.}, $T_{c}$, $T_{w}$, $S_{500_{c}}$, and $S_{500_{w}}$. The uncertainties on \logOH, $\alpha_{\it CO}$, and $[\frac{O}{H}]_{\odot}$ are quoted in dex, and so the perturbations were carried out in logarithmic space; for the other parameters, the uncertainties are stated as simple $\pm$ values, and so the perturbations were carried out in linear space. The uncertainties on $S_{{\it H{\sc I}}}$ and $S_{\it CO}$ were taken to be the Root-Mean-Square (RMS) noise values quoted in \citet{Boselli2014B}. For sources for which \citet{Boselli2014B} do not give an RMS noise, we assume a `worst-case scenario' detection with a Signal-to-Noise Ratio (SNR) of 2.

The re-sampled values of $T_{c}$, $T_{w}$, $S_{500_{c}}$, and $S_{500_{w}}$ were produced by re-fitting the SED for each bootstrapping iteration; the 22--500\,\micron\ fluxes were randomly perturbed according to a Gaussian distribution defined by their uncertainties, and a best fit was then made to this re-sampled SED. Whilst we had used a fixed $\beta = 2$ when carrying out the SED fits to the actual measured fluxes of each source, we left $\beta$ {\it free} when performing the bootstrapping; otherwise our uncertainties would have been artificially small.

The distribution of bootstrapped values of $\kappa_{500}$ for each galaxy are shown in Figure~\ref{Fig:Kappa_Bootstrap_Hist_Grid}. The distributions are generally asymmetric, and as such we define our uncertainties asymmetrically; specifically, by the $66.\dot{6}$\textsuperscript{th} percentile away from the determined value along the bootstrapped distribution, in each direction. The resulting values are given in Table~\ref{Table:Kappa} (also listed for each source is the approximately-equivalent logarithmic uncertainty, in dex, defined by the $66.\dot{6}$\textsuperscript{th} percentile away from the determined value in absolute terms). The median value of $\kappa_{500} = 0.051\,{\rm m^{2}\,kg^{-1}}$ is in agreement with the values of all of the sources to within their individual uncertainties, with the exception of NGC\,4567.

To determine the uncertainty on the average value of $\kappa_{500}$, we merged the bootstrapped distributions of all 22 galaxies in our sample; the resulting combined distribution is plotted in Figure~\ref{Fig:Kappa_Bootstrap_Hist}. We use this distribution to define the uncertainty on the median value of $\kappa_{500}$, as before. With these uncertainties, we therefore find a value of the dust mass absorption coefficient \kappad\ at a wavelength of 500\,\micron\ of $\kappa_{500} = 0.051^{+0.070}_{-0.026}\,{\rm m^{2}\,kg^{-1}}$. The approximately-equivalent logarithmic uncertainty on this value is 0.24\,dex. 

\subsection{Milky Way ${\bf \alpha_{\it CO}}$} \label{Subsection:Milky_Way_Alpha_CO}

For comparison and redundancy, we also determine $\kappa_{500}$ using molecular gas masses determined using a constant Milky Way CO-to-H$_{2}$ conversion factor of $\alpha_{\it CO} = 3.2\, {\rm M_{\odot}}$. This value is considered to be uncertain by a factor of $\sim$2 (see review in \citealp{Saintonge2011A}). Using this value of $\alpha_{\it CO}$, but otherwise proceeding in the exact same manner as before, gives a \kappad\ value of $\kappa_{500} = 0.071^{+0.096}_{-0.036}\,{\rm m^{2}\,kg^{-1}}$. This is well within the uncertainty of the value produced using the metallicity-dependant $\alpha_{\it CO}$.


\section{Variation in \kappad} \label{Section:Variation}

We now examine whether \kappad\ correlates with any of the properties of the galaxies in our sample. The properties we consider are: $M_{\star}$ (stellar mass), $M_{d}$, $M_{d}/M_{\star}$, $T_{c}$, $L_{\it TIR}$ (total infrared luminosity), $M_{{\it H{\sc I}}}$, $M_{H_{2}}$, $M_{{\it H{\sc I}}}/M_{\star}$, $M_{{\it H{\sc I}}}/M_{H_{2}}$, ${\it SFR}$ (star formation rate), ${\it SFR}/M_{\star}$, $Z$, D25 (the angular diameter at the 25\textsuperscript{th} magnitude per square arcsecond isophote), and $L_{\it TIR}/L_{\it FUV}$ (a proxy for FUV dust attenuation). The values of  $M_{\star}$, $L_{\it TIR}$, ${\it SFR}$, D25 and $L_{\it FUV}$ are the same as used in Section~5.1.1 of \citet{CJRClark2015A}.

In turn, we plotted the $\kappa_{500}$ values of the galaxies in our sample against each of the parameters in question, and found the best-fit straight line, using a $\chi^{2}$-minimisation approach. We then bootstrapped these fits 1000 times each, randomly perturbing the value of $\kappa_{500}$ for every galaxy, according to the distributions shown in Figure~\ref{Fig:Kappa_Bootstrap_Hist_Grid}. The results are compatible with there being no correlation of $\kappa_{500}$ with any of the parameters considered. Similarly, for each relationship, Spearman rank correlation tests do not allow us to reject the null hypothesis (of there being no correlation) to a likelihood of \textless\,0.01. We note, however, that the size of our sample is small compared to the scale of the uncertainties on $\kappa_{500}$; hence we in no way rule out that such correlations could exist, but not be detectable in our data.

\section{Comparison to Previous Estimates} \label{Section:Comparison}

As can be seen in Figure~\ref{Fig:Year_vs_Kappa}, our determined value of \kappad\ is low compared to existing literature values (although the very commonly-used values of \citealp{Draine2003A} and \citealp{Draine2007A} are within our bootstrapped uncertainty). Converting the \James\ value to a wavelength of 500\,\micron\ for comparison with our own, as per Equation~\ref{Equation:Kappa_Wavelength}, translates it to $\kappa_{500} = 0.20\pm0.06\,{\rm m^{2}\,kg^{-1}}$; a factor of 3.92 larger than our value, despite the fact we employ the same fundamental technique; it is important to address the reasons for this. 

As discussed in Section~\ref{Section:Method}, our consideration of systematic effects, using $\xi$ (ie, the contribution of helium and metals to a galaxy's gas mass) and $\delta_{O}$ (ie, oxygen-depletion in H{\sc ii} regions) will account for a factor 1.86 reduction in \kappad\ relative to \James. We also employ a slightly larger value of \epsilond\ (0.5 versus 0.456), leading to a further factor 1.10 reduction in \kappad. However, our use of a more modern, smaller value of $f_{Z_{\odot}}$ (0.0134 versus 0.019) works in the opposite direction, diminishing the reduction in \kappad\ by a factor of 0.71. In combination, these global effects correspond to a factor 1.42 reduction in \kappad\ relative to \James; additional effects must be at work to explain the full difference between our values. 

Fortunately, M\,61 (NGC\,4303) is present in both the \James\ sample and our own, allowing for a direct comparison of the reasons for the difference in values for a specific case. For M\,61, we find that our result of $\kappa_{500} = 0.039\,{\rm m^{2}\,kg^{-1}}$ (see Table~\ref{Table:Kappa}) is a factor of 5.13 less than the $\kappa_{500} = 0.20\,{\rm m^{2}\,kg^{-1}}$ value that  \James\ found for their sample as a whole. \James\ did not work out values of \kappad\ for individual sources, but do say that there is scatter of a factor of 2 around their sample-wide relation, hence we need to account for a factor of $5.31^{+5.13}_{-2.57}$ difference between our \kappad\ for M\,61, and the \James\ sample-wide value. In Appendix~\ref{AppendixSection:Deviation}, we work through each of the sources of difference in our values of \kappad\ in the case of M\,61, and find that a factor of 6.31 is to be expected, well within the predicted range -- highlighting the dramatic influence of considering systematic effects and utilising superior observations.

During the \hersc\ Science Demonstration Phase, \citet{Eales2010C} applied a resolved variant of the \James\ method (where they pegged the dust mass surface density to the gas mass surface density, assuming a fixed Schmidt-Kennicutt law and a Milky Way dust-to-gas ratio) to observations of two of the first HRS targets, M\,99 and M\,100. Whilst their analysis was limited by the fact they had \hersc\ observations of only these two galaxies, and assumed an uncorrected solar metallicity, they noted that once the HRS was complete it would represent an ideal dataset for a full determination of \kappad. Converted to our 500\,\micron\ reference wavelength as per Equation~\ref{Equation:Kappa_Wavelength}, they found $\kappa_{500} = 0.027\,{\rm m^{2}\,kg^{-1}}$ for M\,99, and $\kappa_{500} = 0.031\,{\rm m^{2}\,kg^{-1}}$ for M\,100. These values are smaller than (although still within the uncertainties of) the values we find for these galaxies; indeed, as seen in Figure~\ref{Fig:Year_vs_Kappa}, these are the smallest of all values of \kappad\ reported in the literature. The nature of the resolved analysis employed by \citet{Eales2010C} makes it difficult to directly compare our values; however, the fact that they could only consider pixels with significant detections of dust, \HI, and CO, will have limited their analysis to regions with higher ISM surface-density -- regions which might therefore have systematically different dust properties.

In contrast to this, the reader should note that we are using integrated, galaxy-wide values for all of our measurements. This is relevant to the \HI\ component in particular, as it is conceivable that the mass in an extended \HI\ disc may not have a direct bearing upon the properties of the dust and gas found within the inner disc. If so, we would be overestimating the $M_{{\it H{\sc I}}}$ term in Equation~\ref{Equation:Kappa}, and hence finding an artificially small value of \kappad. However, we have reasons to believe that this is unlikely to be significantly distorting our result. The work of \citet{Menard2010A} and Smith et al. ({\it submitted}) indicate that a significant fraction of a galaxy's dust mass is found outside the main stellar disc (\citeauthor{Menard2010A} infer from quasar reddening that half the dust mass is found beyond the stellar discs of galaxies, whilst Smith et al. perform submm stacking of the HRS galaxies and find that 10\% of galaxies' dust mass lies outside the optical D25, following an exponential profile). This shouldn't necessarily be surprising, given that the ISM in the outer discs of massive spiral galaxies has been shown to be metal-rich, even out to large optical radii \citep{Werk2011A,Patterson2012C}. Furthermore, Dunne at al. ({\it in prep.}) show that significant masses of dust can reside in an atomic-gas-dominated medium where there is only a minimal molecular gas component. The HRS submm photometry of \citet{Ciesla2012B} used large apertures, with sizes 1.4--3.3 times the optical D25 for the galaxies in our sample; as such, the (likely faint) emission from any extended dust component will be incorporated into the submm fluxes we use. Similarly, given the observing strategy described by \citep{Boselli2013A}, the HRS drift-scan spectroscopy sampled the region beyond the optical disc for 82\,per\,cent of the galaxies in our sample; in these cases, the metallicity of the outer disc will hence have been sampled\footnote{For those galaxies, an average of 30\,per\,cent of the solid angle scanned by the spectroscopy was beyond the optical D25.} (in a luminosity-weighted manner). Regardless, the galaxies of our sample do not actually appear to posses significantly-extended atomic gas discs. Resolved 21\,cm observations are available in the literature for 59\,per\,cent of our sample\footnote{Resolved 21\,cm data measurements from \citet{Knapen1997B} for NGC\,3631, \citet{Verheijen2001E} for NGC\,3953, and the {\sc Vla} Imaging of Virgo spirals in Atomic gas (VIVA, \citealp{Chung2009C}) survey for the rest. Of the galaxies with resolved 21\,cm data available, 23\,per\,cent are not associated with the Virgo Cluster; for the sample as a whole, the fraction is 32\,per\,cent.}, and for these objects the submm photometric apertures encircle an average of 95\,per\,cent of the detected 21\,cm emission (with a minimum of 88\,per\,cent) -- whist the optical D25 isophotes contain an average of 80\,per\,cent of the 21\,cm emission. As such, even in the unphysically extreme scenario where the atomic gas beyond the optical D25 contains no dust at all,  a value of \kappad\ determined using our method would be underestimated by no more than a factor of 1.2.

Directly comparing our value for \kappad\ to those determined via other means is problematic. At the most fundamental level our method is similar to many theoretical approaches, in that the initial consideration is the fraction of metals depleted into dust grains  -- indeed, our final value of \kappad\ is compatible with that determined by \citet{Draine2003A} and \citet{Draine2007A}, who work from that very premise. However, beyond this first step, it is very difficult to compare an empirical result such as ours to theoretical values that arise from considerations of astrochemistry and complex Mie theory calculations. Similar calculations go into estimates based upon radiative transfer modelling, which additionally rely upon assumptions about the optical properties and/or geometry of the dust being modelled, making them equally troublesome to compare to. However, whilst these differences make other values for \kappad\ impractical to compare to our own, we argue that this renders them highly complementary.

\section{Conclusions} \label{Section:Conclusions}

We apply the method of \citet{James2002} to the rich, homogeneous dataset of the HRS. This technique enables us to determine the dust mass in a galaxy without needing to use FIR--submm photometry; by relating this calculated dust mass to the observed dust emission, we can empirically find the value of the dust mass absorption coefficient, \kappad. 

We find a value of \kappad\ at a wavelength of 500\,\micron\ of $\kappa_{500} = 0.051^{+0.070}_{-0.026}\,{\rm m^{2}\,kg^{-1}}$.  The uncertainty on this value was determined rigorously, by bootstrapping for every input parameter, taking the uncertainty on the prescriptions employed, and the individual measurements used.

Empirical determinations of \kappad, such as this, provide a vital counterpoint to the development of theoretical dust models, as there are precious few ways in which the properties of dust can be observationally determined.


We note that our value for \kappad\ is susceptible to an additional systematic offset of up to 0.7\,dex, due to the uncertainty in the absolute metallicity scale calibration \citep{Kewley2008C}. However, even when the full worst-case offset of 0.7\,dex is combined with the results of our bootstrapping analysis, the combined uncertainty of $\sim$\,0.74\,dex is {\it still} less than the generally assumed `order-of-magnitude' uncertainty on \kappad. 

With a single exception, the values of \kappad\ we determine for the galaxies in our sample agree to within their uncertainties. Na\"ively speaking, one would expect this to be true for only $\sim$\,66\,per\,cent of the sample. This suggests that the uncertainty values (the majority of which we take from the literature, with the exception of those derived from our SED fits) incorporated into our bootstrapping analysis have, on average, been overestimated by their respective authors.

We find no evidence that \kappad\ varies as a function of any of the properties of the galaxies in our sample. However, as we opted to limit our sample to HRS galaxies that share the full set of homogeneous integrated measurements required, our sample consists only of relatively massive ($10^{9.7}<{M_{\star}}<10^{11.0}\,{\rm M_{\odot}}$), high metallicity (8.61\,\textless\,\logOH\,\textless\,8.86) systems. To truly establish if (and how) the value of \kappad\ changes between galaxies is vital for the field -- despite the huge advances in dust astrophysics over the past 10--15 years, we still have little idea if and how the value of \kappad\ differs between galaxies. But doing so requires homogeneous high-quality data for a wider range of systems than are available at present.

Fortunately, large Integral Field Unit (IFU) spectroscopy surveys such as CALIFA\footnote{Calar Alto Legacy Integral Field spectroscopy Area survey \citep{Sanchez2012B}.}, SAMI\footnote{Sydney-{\sc Aao} Multi-object Integral-field spectrograph \citep{Bryant2015B}.}, and MaNGA\footnote{Mapping Nearby Galaxies at {\sc Apo} \citep{Bundy2015B}.} mean that integrated metallicity measurements will soon be available for vastly greater numbers of galaxies. Similarly, statistically-large local-Universe CO surveys, such as JINGLE\footnote{{\sc Jcmt} dust and gas In Nearby Galaxies Legacy Exploration.}, are becoming more common. Once these datasets become available, it will be possible to use the method we employ here to constrain the value of \kappad\ for galaxies with a far broader range of properties.

\section{Acknowledgements}

The authors would like to thank the anonymous referee for their valuable comments, which lead to substantive improvements to this work.

CJRC acknowledges support from the European Research Council (ERC) in the form of the FP7 project DustPedia (PI Jon Davies, proposal 606824) and the Science and Technology Facilities Council (STFC) Doctoral Training Grant scheme, and kindly thanks Steve Eales and Matt Smith for helpful conversations. 

SPS acknowledges support from the STFC Doctoral Training Grant scheme. JID acknowledges support from the ERC in the form of the FP7 project DustPedia (proposal 606824). HLG acknowledges support from the ERC in the form of Consolidator Grant project {\sc CosmicDust} (proposal 647939).

This research has made use of Astropy\footnote{\url{http://www.astropy.org/}}, a community-developed core Python package for Astronomy \citep{Astropy2013}. This research has made use of TOPCAT\footnote{\url{http://www.star.bris.ac.uk/~mbt/topcat/}} \citep{Taylor2005A}, which was initially developed under the UK Starlink project, and has since been supported by PPARC, the VOTech project, the AstroGrid project, the AIDA project, the STFC, the GAVO project, the European Space Agency, and the GENIUS project. This research has made use of NumPy\footnote{\url{http://www.numpy.org/}} \citep{Walt2011B}, SciPy\footnote{\url{http://www.scipy.org/}}, and MatPlotLib\footnote{\url{http://matplotlib.org/}} \citep{Hunter2007A}. This research has made use of the SIMBAD\footnote{\url{http://simbad.u-strasbg.fr/simbad/}} database \citep{Wenger2000D} and the VizieR\footnote{\url{http://vizier.u-strasbg.fr/viz-bin/VizieR}} catalogue access tool \citep{Ochsenbein2000B}, both operated at CDS, Strasbourg, France. This research has made use of the NASA/IPAC Infrared Science Archive (IRSA\footnote{\url{http://irsa.ipac.caltech.edu/frontpage/}}), operated by the Jet Propulsion Laboratory, California Institute of Technology, under contract with the National Aeronautics and Space Administration. This research has made use of code written by Adam Ginsburg\footnote{\url{https://github.com/keflavich}}, kindly made available under the GNU General Public License\footnote{\url{http://www.gnu.org/copyleft/gpl.html}}.

\hersc\ is an ESA space observatory with science instruments provided by European-led Principal Investigator consortia and with important participation from NASA. The \hersc\ spacecraft was designed, built, tested, and launched under a contract to ESA managed by the \hersc/\planck\ Project team by an industrial consortium under the overall responsibility of the prime contractor Thales Alenia Space (Cannes), and including Astrium (Friedrichshafen) responsible for the payload module and for system testing at spacecraft level, Thales Alenia Space (Turin) responsible for the service module, and Astrium (Toulouse) responsible for the telescope, with in excess of a hundred subcontractors.

\def\ref@jnl#1{{\rmfamily #1}}%
\newcommand\aj{\ref@jnl{AJ}}%
\newcommand\araa{\ref@jnl{ARA\&A}}%
\newcommand\apj{\ref@jnl{ApJ}}%
\newcommand\apjl{\ref@jnl{ApJ}}%
\newcommand\apjs{\ref@jnl{ApJS}}%
\newcommand\ao{\ref@jnl{Appl.~Opt.}}%
\newcommand\apss{\ref@jnl{Ap\&SS}}%
\newcommand\aap{\ref@jnl{A\&A}}%
\newcommand\aapr{\ref@jnl{A\&A~Rev.}}%
\newcommand\aaps{\ref@jnl{A\&AS}}%
\newcommand\azh{\ref@jnl{AZh}}%
\newcommand\baas{\ref@jnl{BAAS}}%
\newcommand\jrasc{\ref@jnl{JRASC}}%
\newcommand\memras{\ref@jnl{MmRAS}}%
\newcommand\mnras{\ref@jnl{MNRAS}}%
\newcommand\pra{\ref@jnl{Phys.~Rev.~A}}%
\newcommand\prb{\ref@jnl{Phys.~Rev.~B}}%
\newcommand\prc{\ref@jnl{Phys.~Rev.~C}}%
\newcommand\prd{\ref@jnl{Phys.~Rev.~D}}%
\newcommand\pre{\ref@jnl{Phys.~Rev.~E}}%
\newcommand\prl{\ref@jnl{Phys.~Rev.~Lett.}}%
\newcommand\pasp{\ref@jnl{PASP}}%
\newcommand\pasj{\ref@jnl{PASJ}}%
\newcommand\qjras{\ref@jnl{QJRAS}}%
\newcommand\skytel{\ref@jnl{S\&T}}%
\newcommand\solphys{\ref@jnl{Sol.~Phys.}}%
\newcommand\sovast{\ref@jnl{Soviet~Ast.}}%
\newcommand\ssr{\ref@jnl{Space~Sci.~Rev.}}%
\newcommand\zap{\ref@jnl{ZAp}}%
\newcommand\nat{\ref@jnl{Nature}}%
\newcommand\iaucirc{\ref@jnl{IAU~Circ.}}%
\newcommand\aplett{\ref@jnl{Astrophys.~Lett.}}%
\newcommand\apspr{\ref@jnl{Astrophys.~Space~Phys.~Res.}}%
\newcommand\bain{\ref@jnl{Bull.~Astron.~Inst.~Netherlands}}%
\newcommand\fcp{\ref@jnl{Fund.~Cosmic~Phys.}}%
\newcommand\gca{\ref@jnl{Geochim.~Cosmochim.~Acta}}%
\newcommand\grl{\ref@jnl{Geophys.~Res.~Lett.}}%
\newcommand\jcp{\ref@jnl{J.~Chem.~Phys.}}%
\newcommand\jgr{\ref@jnl{J.~Geophys.~Res.}}%
\newcommand\jqsrt{\ref@jnl{J.~Quant.~Spec.~Radiat.~Transf.}}%
\newcommand\memsai{\ref@jnl{Mem.~Soc.~Astron.~Italiana}}%
\newcommand\nphysa{\ref@jnl{Nucl.~Phys.~A}}%
\newcommand\physrep{\ref@jnl{Phys.~Rep.}}%
\newcommand\physscr{\ref@jnl{Phys.~Scr}}%
\newcommand\planss{\ref@jnl{Planet.~Space~Sci.}}%
\newcommand\procspie{\ref@jnl{Proc.~SPIE}}%
\newcommand\jcap{\ref@jnl{JCAP}}%
\bibliographystyle{mn2e}
\bibliography{ChrisBib}

\newpage

\appendix

\section{Deviation from James et al. in the Case of M\,61} \label{AppendixSection:Deviation}

As described in Sections~2 and 7, we find that our calculated values of \kappad\ should be expected to be smaller than those of \James\ by a factor of 1.42, due to global effects. Here we describe in detail the further differences that should be expected to arise in the useful case of M\,61, which is in both our sample, and that of \James.

Whereas \James\ quote a metallicity of $[ 12 + {\rm log}_{10} \frac{O}{H} ] = 9.01$ for M\,61, taken from spectra of individual H{\sc ii} regions\footnote{Spectra taken by \citet{Shields1991C} and \citet{Henry1992A}, and compiled by \citet{Zaritsky1994C}.}, we use a metallicity of $[ 12 + {\rm log}_{10} \frac{O}{H} ] = 8.67$, derived from integrated drift-scan spectroscopy \citep{Hughes2013B,Boselli2013A} -- which should therefore be the superior measurement. This difference in metallicity contributes a further factor 2.19 to the reduction in our value in \kappad\ relative to that of \James. However, the difference in metallicity also gives us a smaller $\alpha_{\it CO}$, and hence gas mass, by a factor of 0.71.

\James\ used a lower value of $S_{\it CO}$ for M\,61 than we do. Their value is is derived from a series of 7 pointings\footnote{Observations made by \citet{Kenney1988AC}, and complied by \citet{Young1995A}} along the major axis of M\,61, extrapolated to the rest of the disc (rendering them unable to account for azimuthal variation) -- whereas the value of $S_{\it CO}$ we take from \citet{Boselli2014B} is derived from a mapping of all detected CO emission, with no reliance upon extrapolation, and so is presumably a far more accurate value.

\James\ also quoted a much smaller \HI\ mass for M\,61 than the one we take from \citet{Boselli2014B}, leading to an additional factor 2.19 difference in our expected value of \kappad. Unfortunately we are unable to definitively address the underlying reasons for this discrepancy, as \James\ do not seem to provide a reference for the \HI\ mass they use. We note however their \HI\ mass is the same as that given by \citealp{Magrini2011A}, who integrated over the \HI\ radial profile produced by \citealp{Skillman1996E} out to 0.7 optical radii (due to signal-to-noise constraints in their ancillary data). However, it is unclear if \James\ intentionally chose an \HI\ value that only extends out to some fraction of the optical radius of M\,61, as most of the \HI\ masses employed by \James\ are those compiled by the {\sc Scuba} Local Universe Galaxy Survey (SLUGS, \citealp{Dunne2000A}), whose 21\,cm values are point-source integrated measurements taken from the literature. The \HI\ mass we take from \citet{Boselli2014B} comes from the 84.71\,Jy flux measured by the ALFALFA survey (Arecibo Legacy Fast {\sc ALFA}, \citealp{Giovanelli2005D,Haynes2011}), which is in excellent agreement with the 85.2\,Jy flux measured at the Westerbork Synthesis Radio Telescope by \citet{Popping2011B}, and the 85.1\,Jy flux measured at the Parkes Radio Telescope by \citet{Popping2011F}. As such, it seems that the measurement we use is accurate; possibly simply a benefit of more modern observations.

M\,61 also illustrates the effect that superior data from \hersc\ has on SED-fitting. The $\sim$\,0.9 order-of-magnitude gap in wavelength coverage suffered by \James\ between the IRAS 100\,\micron\ and SCUBA 850\,\micron\ points means that they had no coverage of the dust emission peak, making it difficult to constrain dust temperatures. As a result they fixed the temperature of their cold dust component to 20\,K when performing their SED fitting. The general shapes of their SED fits are very different from our own (compare their Figure~2 to our Figure~B1), with a much more prominent warm component, clearly incompatible with the \hersc\ photometry. Whereas they find that M\,61 has a cold-to-warm dust mass ratio of $M_{c}/M_{w} = 83$ (see their Table~2), we find $M_{c}/M_{w} = 7,500$ (their sample median cold-to-warm dust mass ratio is only 19, compared to our median of 2,310). In the case of M\,61, the net effect is an increase in the value of the section of Equation~10 that incorporates the SED parameters\footnote{Specifically, the term: $\left( \frac{ S_{\lambda_{w}} }{ B_{\lambda}(T_{w}) } + \frac{ S_{\lambda_{c}} }{ B_{\lambda}(T_{c}) } \right)$.}, which decreases by a factor of 0.87 the expected deviation between our value of \kappad\ and that of \James.

Combining all of these effects leads us to expect our value of \kappad\ for M\,61 should be a factor of 6.31 smaller than that of \James\ (a factor of 1.41 from $\xi$, 1.32 from $\delta_{O}$, 1.10 from \epsilond, 0.71 from $f_{Z_{\odot}}$, 2.19 from $M_{{\it H{\sc I}}}$, 1.47 from $S_{\it CO}$, 2.19 from $Z$, 0.71 from $\alpha_{\it CO}$, and 0.87 from SED-fitting). The individual effects contributing to the overall difference are all attributable to our consideration of systematic effects, or our use of superior observational measurements (for \HI, CO, metallicity, and FIR-submm data).

\section{Properties of the Sample Galaxies} \label{AppendixSection:Tables}

Here we present plots and tables detailing the properties of the galaxies in our sample. 

\begin{table*}
\begin{center}
\caption{Basic properties of the HRS galaxies that we study in this work. Values taken from \citet{Boselli2010} and \citet{Cortese2012A}.} 
\label{AppendixTable:Sample}
\begin{tabular}{lrrrrrr}
\toprule \toprule
\multicolumn{1}{c}{Name} &
\multicolumn{1}{c}{RA} &
\multicolumn{1}{c}{Dec} &
\multicolumn{1}{c}{Distance} &
\multicolumn{1}{c}{Heliocentric Velocity} &
\multicolumn{1}{c}{Morphology} &
\multicolumn{1}{c}{Stellar Mass} \\
\multicolumn{1}{c}{} &
\multicolumn{1}{c}{(J2000 deg)} &
\multicolumn{1}{c}{(J2000 deg)} &
\multicolumn{1}{c}{(Mpc)} &
\multicolumn{1}{c}{(km s$^{-1}$)} &
\multicolumn{1}{c}{(Hubble stage)} &
\multicolumn{1}{c}{(${\rm log_{10}\, M_{\odot}}$)} \\
\midrule
NGC\,3437 & 163.149 & 22.934 & 18.2 & 1277 & 7 & 9.69\\ 
NGC\,3631 & 170.262 & 53.170 & 16.5 & 1155 & 7 & 9.92\\ 
NGC\,3683 & 171.883 & 56.877 & 24.4 & 1708 & 7 & 10.20\\ 
NGC\,3953 & 178.454 & 52.327 & 15.0 & 1050 & 6 & 10.60\\ 
NGC\,4030 & 180.098 & -1.100 & 20.8 & 1458 & 6 & 10.54\\ 
M\,98 & 183.451 & 14.900 & 17.0 & -135 & 4 & 10.65\\ 
NGC\,4212 & 183.914 & 13.902 & 17.0 & -83 & 7 & 10.01\\ 
M\,99 & 184.707 & 14.416 & 17.0 & 2405 & 7 & 10.39\\ 
M\,61 & 185.479 & 4.474 & 17.0 & 1568 & 6 & 10.51\\ 
M\,100 & 185.729 & 15.822 & 17.0 & 1575 & 6 & 10.71\\ 
M\,86 & 186.531 & 13.113 & 17.0 & 234 & 5 & 10.04\\ 
M\,88 & 187.997 & 14.420 & 17.0 & 2284 & 5 & 10.98\\ 
NGC\,4527 & 188.535 & 2.654 & 17.0 & 1736 & 6 & 10.67\\ 
NGC\,4535 & 188.585 & 8.198 & 17.0 & 1962 & 7 & 10.45\\ 
NGC\,4536 & 188.613 & 2.188 & 17.0 & 1807 & 6 & 10.26\\ 
NGC\,4567 & 189.136 & 11.258 & 17.0 & 2277 & 6 & 9.92\\ 
NGC\,4568 & 189.143 & 11.239 & 17.0 & 2255 & 6 & 10.33\\ 
M\,60 & 190.885 & 11.583 & 17.0 & 1422 & 7 & 10.19\\ 
NGC\,4651 & 190.928 & 16.393 & 17.0 & 797 & 7 & 10.13\\ 
NGC\,4654 & 190.986 & 13.127 & 17.0 & 1039 & 8 & 10.14\\ 
NGC\,4689 & 191.940 & 13.763 & 17.0 & 1620 & 6 & 10.19\\ 
NGC\,5248 & 204.384 & 8.885 & 16.5 & 1152 & 6 & 10.43\\ 
\bottomrule
\end{tabular}
\end{center}
\end{table*}

\begin{table*}
\begin{center}
\caption{Dust properties of the HRS galaxies that we study in this work. The temperatures and 500\,\micron\ fluxes of the cold and warm dust components were determined by fitting a two-component modified blackbody SED to the published HRS photometry, as detailed in Section~4. Note that the quoted uncertainties are merely representative; when bootstrapping to find the total uncertainty in \kappad, we re-fit a bootstrapped SED for every iteration (as described in Section~5).} 
\label{AppendixTable:Dust}
\begin{tabular}{lrrrrrrrr}
\toprule \toprule
\multicolumn{1}{c}{Name} &
\multicolumn{1}{c}{$T_{c}$} &
\multicolumn{1}{c}{$\Delta\,T_{c}$} &
\multicolumn{1}{c}{$T_{w}$} &
\multicolumn{1}{c}{$\Delta\,T_{w}$} &
\multicolumn{1}{c}{$S_{500_{c}}$} &
\multicolumn{1}{c}{$\Delta\,S_{500_{c}}$} &
\multicolumn{1}{c}{$S_{500_{w}}$} &
\multicolumn{1}{c}{$\Delta\,S_{500_{w}}$} \\
\cmidrule(lr){2-5}
\multicolumn{1}{c}{} &
\multicolumn{4}{c}{(K)} &
\multicolumn{1}{c}{(Jy)} &
\multicolumn{1}{c}{(dex)} &
\multicolumn{1}{c}{(mJy)} &
\multicolumn{1}{c}{(dex)} \\
\midrule
NGC\,3437 & 23.26 & 4.15 & 50.60 & 15.95 & 1.25 & 0.17 & 27.44 & 1.28\\ 
NGC\,3631 & 20.43 & 4.27 & 35.50 & 19.05 & 3.17 & 0.33 & 74.84 & 1.69\\ 
NGC\,3683 & 24.16 & 3.47 & 63.47 & 14.29 & 1.53 & 0.21 & 9.73 & 0.99\\ 
NGC\,3953 & 18.92 & 1.47 & 46.18 & 18.98 & 5.09 & 0.50 & 5.56 & 1.58\\ 
NGC\,4030 & 21.83 & 2.41 & 68.45 & 17.33 & 5.01 & 0.36 & 8.38 & 1.41\\ 
M\,98 & 19.01 & 1.82 & 41.91 & 18.18 & 4.72 & 0.29 & 25.75 & 1.34\\ 
NGC\,4212 & 21.65 & 3.70 & 66.46 & 18.19 & 1.78 & 0.25 & 4.29 & 1.19\\ 
M\,99 & 22.44 & 1.43 & 72.43 & 17.75 & 8.68 & 0.35 & 3.12 & 1.78\\ 
M\,61 & 22.48 & 1.64 & 68.54 & 17.35 & 8.09 & 0.35 & 5.36 & 2.06\\ 
M\,100 & 20.76 & 1.33 & 59.21 & 18.53 & 9.74 & 0.49 & 3.55 & 1.91\\ 
M\,86 & 20.82 & 2.42 & 67.53 & 16.96 & 2.01 & 0.31 & 3.18 & 1.10\\ 
M\,88 & 20.65 & 1.48 & 58.84 & 18.42 & 8.46 & 0.47 & 6.29 & 1.62\\ 
NGC\,4527 & 21.52 & 3.04 & 59.97 & 18.04 & 6.75 & 0.30 & 29.84 & 1.23\\ 
NGC\,4535 & 19.25 & 1.79 & 51.09 & 19.24 & 5.88 & 0.45 & 11.66 & 1.60\\ 
NGC\,4536 & 17.93 & 5.18 & 31.89 & 18.10 & 3.96 & 0.21 & 924.81 & 1.24\\ 
NGC\,4567 & 19.66 & 1.68 & 50.84 & 2.51 & 1.32 & 0.07 & 41.26 & 0.25\\ 
NGC\,4568 & 22.50 & 3.78 & 72.36 & 15.43 & 3.81 & 0.30 & 4.26 & 1.15\\ 
M\,60 & 21.49 & 3.06 & 67.50 & 17.01 & 1.61 & 0.26 & 3.16 & 1.07\\ 
NGC\,4651 & 18.44 & 3.96 & 27.64 & 19.05 & 1.79 & 0.29 & 337.32 & 1.57\\ 
NGC\,4654 & 20.60 & 2.16 & 41.77 & 17.38 & 4.48 & 0.35 & 57.69 & 1.48\\ 
NGC\,4689 & 19.88 & 4.27 & 54.47 & 17.39 & 1.73 & 0.36 & 2.29 & 1.63\\ 
NGC\,5248 & 21.10 & 2.21 & 53.23 & 17.64 & 5.89 & 0.33 & 27.66 & 1.43\\ 
\bottomrule
\end{tabular}
\end{center}
\end{table*}

\begin{table*}
\begin{center}
\caption{Gas properties of the HRS galaxies that we study in this work. \HI\ and CO measurements taken from \citet{Boselli2014B}. Gas-phase metallicity values taken from \citet{Hughes2013B}, which used the spectra of \citet{Boselli2013A}. Note that the metallicities given in the $Z$ column have been corrected for gas-phase oxygen depletion by a factor of $\delta_{O} = 1.32$, as per Equation~3; as a result, they represent ISM metallicities, not gas-phase metallicities.} 
\label{AppendixTable:Gas}
\begin{tabular}{lrrrrrrrr}
\toprule \toprule
\multicolumn{1}{c}{Name} &
\multicolumn{1}{c}{$S_{{\it H{\sc I}}}$} &
\multicolumn{1}{c}{$\Delta\,S_{{\it H{\sc I}}}$} &
\multicolumn{1}{c}{$S_{\it CO}$} &
\multicolumn{1}{c}{$\Delta\,S_{\it CO}$} &
\multicolumn{1}{c}{\logOH} &
\multicolumn{1}{c}{$\Delta\,[ 12 + {\rm log}_{10} \frac{O}{H} ]$} &
\multicolumn{1}{c}{$Z$} &
\multicolumn{1}{c}{$\alpha_{\it CO}$} \\
\cmidrule(lr){2-5}
\multicolumn{1}{c}{} &
\multicolumn{4}{c}{(Jy\,km\,s$^{-1}$)} &
\multicolumn{1}{c}{} &
\multicolumn{1}{c}{(dex)} &
\multicolumn{1}{c}{($Z_{\odot}$)} &
\multicolumn{1}{c}{(M$_{\odot}$\,K$^{-1}$\,km$^{-1}$\,s\,pc$^{-2}$)} \\
\midrule
NGC\,3437 & 22.24 & 2.86 & 190.00 & 90.30 & 8.67 & 0.03 & 1.26 & 9.19\\ 
NGC\,3631 & 50.70 & 25.35$^{a}$ & 1093.00 & 131.40 & 8.64 & 0.17 & 1.18 & 10.55\\ 
NGC\,3683 & 8.27 & 3.06 & 390.00 & 185.30 & 8.67 & 0.10 & 1.26 & 9.19\\ 
NGC\,3953 & 47.79 & 10.28 & 1790.00 & 850.30 & 8.86 & 0.22 & 1.95 & 3.83\\ 
NGC\,4030 & 64.15 & 5.77 & 1050.00 & 498.80 & 8.69 & 0.10 & 1.32 & 8.38\\ 
M\,98 & 74.14 & 2.25 & 940.00 & 446.50 & 8.76 & 0.10 & 1.55 & 6.07\\ 
NGC\,4212 & 13.92 & 2.20 & 491.80 & 59.10 & 8.71 & 0.10 & 1.38 & 7.64\\ 
M\,99 & 77.05 & 2.42 & 4033.00 & 484.80 & 8.73 & 0.12 & 1.45 & 6.97\\ 
M\,61 & 84.71 & 3.14 & 3344.00 & 402.00 & 8.76 & 0.11 & 1.55 & 6.07\\ 
M\,100 & 48.86 & 2.90 & 3148.00 & 378.40 & 8.75 & 0.10 & 1.52 & 6.35\\ 
M\,86 & 7.73 & 2.62 & 786.90 & 94.60 & 8.68 & 0.10 & 1.29 & 8.77\\ 
M\,88 & 29.10 & 2.28 & 2951.00 & 354.70 & 8.77 & 0.10 & 1.59 & 5.80\\ 
NGC\,4527 & 108.50 & 6.50 & 1862.00 & 768.70 & 8.81 & 0.10 & 1.74 & 4.82\\ 
NGC\,4535 & 71.66 & 2.96 & 1377.00 & 165.50 & 8.77 & 0.10 & 1.59 & 5.80\\ 
NGC\,4536 & 74.90 & 15.00 & 1082.00 & 130.10 & 8.70 & 0.21 & 1.35 & 8.00\\ 
NGC\,4567 & 15.64 & 0.36 & 2229.00 & 920.30 & 8.65 & 0.10 & 1.20 & 10.07\\ 
NGC\,4568 & 25.11 & 0.36 & 1050.00 & 498.80 & 8.77 & 0.22 & 1.59 & 5.80\\ 
M\,60 & 7.86 & 2.94 & 881.20 & 363.80 & 8.61 & 0.10 & 1.10 & 12.11\\ 
NGC\,4651 & 62.99 & 15.24 & 350.00 & 166.30 & 8.75 & 0.07 & 1.52 & 6.35\\ 
NGC\,4654 & 50.59 & 2.52 & 1574.00 & 189.20 & 8.65 & 0.07 & 1.20 & 10.07\\ 
NGC\,4689 & 8.36 & 2.09 & 786.90 & 94.60 & 8.66 & 0.10 & 1.23 & 9.62\\ 
NGC\,5248 & 73.58 & 3.09 & 2425.00 & 291.50 & 8.81 & 0.06 & 1.74 & 4.82\\ 
\bottomrule
\end{tabular}
\end{center}
\begin{list}{}{}
\item[$^{a}$] \citet{Boselli2014B} do note quote an RMS value for the 21\,cm flux measurement of NGC\,3631; for this source, we assume a `worst-case scenario' detection with SNR = 2. 
\end{list}
\end{table*}

\begin{figure*}
\begin{center}
\includegraphics[width=0.925\textwidth]{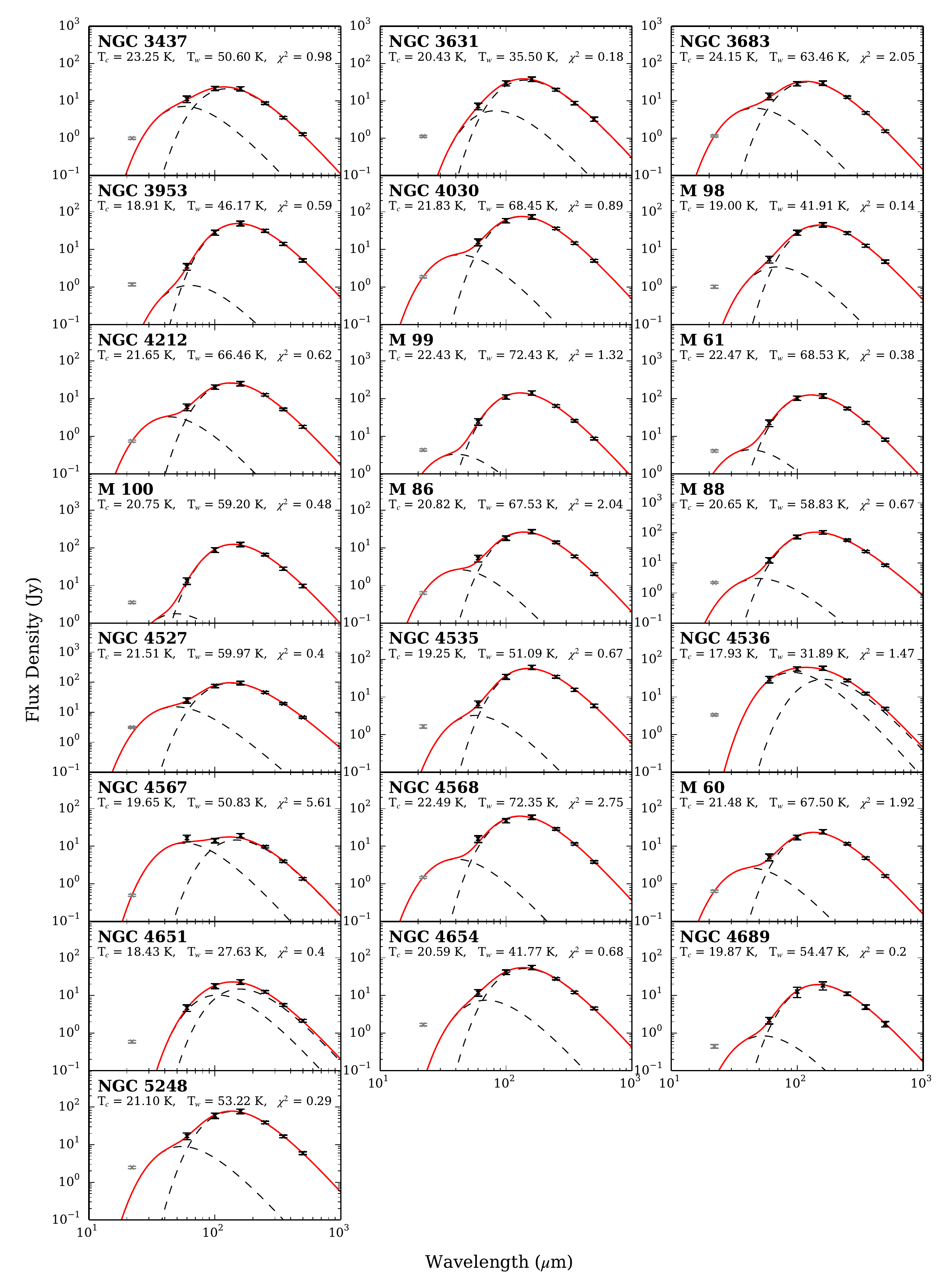}
\caption{Best-fit FIR--submm SEDs of the galaxies in our sample. The two-temperature modified blackbody fits are shown in red, with the contributions from the warm and cold dust components shown by the dashed curves. The grey 22\,\micron\ point was treated as an upper limit.}
\label{AppendixFig:SED_Grid}
\end{center}
\end{figure*}

\end{document}